\documentclass{ieeetj}
\usepackage{cite}
\usepackage{amsmath,amssymb,amsfonts}
\usepackage{algorithmic}
\usepackage{graphicx,color}
\usepackage{textcomp}
\usepackage{xcolor}
\usepackage{hyperref}
\hypersetup{hidelinks=true}
\usepackage{algorithm,algorithmic}
\def\BibTeX{{\rm B\kern-.05em{\sc i\kern-.025em b}\kern-.08em
    T\kern-.1667em\lower.7ex\hbox{E}\kern-.125emX}}
\AtBeginDocument{\definecolor{tmlcncolor}{cmyk}{0.93,0.59,0.15,0.02}\definecolor{NavyBlue}{RGB}{0,86,125}}

\newcommand\blfootnote[1]{%
  \begingroup
  \renewcommand\thefootnote{}\footnote{#1}%
  \addtocounter{footnote}{-1}%
  \endgroup
}

\def\authorrefmark#1{\ensuremath{^{\textbf{#1}}}}

\begin{document}
% \receiveddate{XX Month, XXXX}
% \reviseddate{XX Month, XXXX}
% \accepteddate{XX Month, XXXX}
% \publisheddate{XX Month, XXXX}
% \currentdate{XX Month, XXXX}
% \doiinfo{XXXX.2022.1234567}

\markboth{}{Author {et al.}}

\title{Text-to-Speech for Unseen Speakers via Low-Complexity Discrete Unit–Based Frame Selection}

\author{ISMAIL RASIM ULGEN\authorrefmark{1,*}, SHREERAM SURESH CHANDRA\authorrefmark{1,*}, JUNCHEN LU\authorrefmark{2} \\ BERRAK SISMAN\authorrefmark{1}}
\affil{Center for Language and Speech Processing (CLSP), Johns Hopkins University, Baltimore, MD, USA}
\affil{National University of Singapore, Singapore}
\authornote{This work is supported by NSF CAREER award IIS-2338979.\\
* \textit{equal contribution}}

\begin{abstract}
Synthesizing the voices of unseen speakers remains a persisting challenge in multi-speaker text-to-speech (TTS). Existing methods model speaker characteristics through speaker conditioning during training, leading to increased model complexity and limiting reproducibility and accessibility. A low-complexity alternative would broaden the reach of speech synthesis research, particularly in settings with limited computational and data resources. To this end, we propose SelectTTS, a simple and effective alternative. SelectTTS selects appropriate frames from the target speaker and decodes them using frame-level self-supervised learning (SSL) features. We demonstrate that this approach can effectively capture speaker characteristics for unseen speakers and achieves performance comparable to state-of-the-art multi-speaker TTS frameworks on both objective and subjective metrics. By directly selecting frames from the target speaker's speech, SelectTTS enables generalization to unseen speakers with significantly lower model complexity. Experimental results show that the proposed approach achieves performance comparable to state-of-the-art systems such as XTTS-v2 and VALL-E, while requiring over $8\times$ fewer parameters and $270\times$ less training data. Moreover, it demonstrates that frame selection with SSL features offers an efficient path to low-complexity, high-quality multi-speaker TTS. 

\end{abstract}



\maketitle

\section{INTRODUCTION}
\IEEEPARstart{R}{ecent} advances in text-to-speech (TTS) have shown that with sufficient data and model capacity, neural TTS systems can generate speech of remarkable naturalness and quality \cite{le2024voicebox,chen2025neural,casanova2024xtts}. While large-scale TTS models \cite{lajszczak2024base} brings significant advantages, they also raise concerns of reproducibility, accessibility, and practicality, particularly in multi-speaker TTS, where reproducing speaker timbre for unseen speakers remains a key challenge. 
\blfootnote{\textbf{Speech Samples:}  \url{https://kodhandarama.github.io/selectTTSdemo/}}
\blfootnote{Codes and pre-trained models will be released upon acceptance.}

Multi-speaker TTS for unseen speakers requires solving two problems simultaneously: predicting speech content from text while accurately modeling speaker timbre and acoustics \cite{chen20r_interspeech}. Early approaches addressed this by conditioning models on speaker labels or embeddings \cite{jia2018transfer,casanova2022yourtts, melechovsky2024dart}, enabling speaker-specific control \cite{gibiansky2017deep,cooper2020zero,jawaid2024style}. More recently, zero-shot TTS frameworks have emerged, leveraging neural codec language models \cite{chen2025neural,xin2024rall} and acoustic prompting to achieve speaker generalization. Despite their success, these embedding- and prompting-based methods all rely on conditional speaker modeling, which comes at the cost of substantial model complexity, capacity, and data requirements~\cite{peng2024voicecraft,du2024cosyvoice, wang2024speechx}. Training and deploying such systems demands large-scale resources, limiting reproducibility and hindering broader research and application.

To overcome these limitations, we propose SelectTTS, a simple and effective alternative that eliminates conditional speaker modeling in favor of non-parametric frame selection from self-supervised learning (SSL) features. Rather than tasking the model with generating speaker timbre, SelectTTS directly selects frame-level features from reference speech of the target speaker, enabling lightweight yet accurate synthesis. This design is made possible by SSL speech models~\cite{baevski2020wav2vec,chen2022wavlm}, which learn rich frame-level representations that capture linguistic, speaker, and prosodic information \cite{pasad2021layer}. Their masked prediction objectives \cite{hsu2021hubert} preserve speaker identity while providing semantic robustness. Inspired by kNN-VC \cite{baas23_interspeech}, which demonstrated the effectiveness of frame selection for voice conversion, SelectTTS brings this idea into multi-speaker TTS
% by splitting the problem into two stages:
by proposing a two-stage strategy:
1) Semantic prediction -- predicting frame-level discrete units from text to represent speech content; and 2) Speaker modeling -- selecting intermediate SSL features with the necessary speaker-acoustic information from a reference utterance of the target speaker based on the predicted semantic units. With SelectTTS, we introduce a new paradigm of frame selection-based multi-speaker TTS that directly utilizes frames from unseen target speakers to reproduce their voice. 

The main contributions of this work are as follows:
\vspace{-2mm}
\begin{enumerate}
    \item We propose a multi-speaker TTS strategy that completely separates and simplifies the tasks of semantic prediction and speaker modeling in TTS, making the overall framework easily reproducible and open to further development;
    \item We introduce novel frame selection algorithms, subsequence matching and inverse k-means sampling, that directly select frames from the target speaker to accurately reproduce speaker timbre;
    \item We demonstrate the benefits of leveraging both discrete SSL features (for semantic prediction and frame selection) and continuous SSL features (for vocoding), achieving competitive performance with far lower complexity than large-scale baselines.
\end{enumerate}

\vspace{-1mm}
The remainder of this paper is organized as follows. Section II reviews the related work on multi-speaker TTS, concatenative and unit selection approaches, and the use of SSL features in speech synthesis. Section III introduces the proposed SelectTTS framework in detail. Section IV describes the experimental setup, including implementation details, baselines, and datasets. 
Section V presents objective and subjective evaluation results. Section VI provides discussion. Finally, Section VII concludes the paper. 
%\sr{
\section{RELATED WORK}

\subsection{MODELING UNSEEN SPEAKERS IN MULTI-SPEAKER TTS}

Zero-shot TTS aims to reproduce the vocal characteristics of previously unseen speakers from only a short reference audio. In the deep-learning era, most approaches achieve this by conditioning neural networks on speaker information. A common line of work uses pre-trained speaker embeddings, often derived from speaker verification models, to guide the synthesis \cite{jia2018transfer,cooper2020zero,casanova2021sc}. Another strategy adapts models to new speakers via lightweight fine-tuning or meta-learning~\cite{huang2022meta}, as in UnitSpeech \cite{kim23k_interspeech} and HierSpeech \cite{lee2022hierspeech}. Inspired by large language models, recent systems perform in-context prompting with neural codec language models (e.g., VALL-E \cite{chen2025neural}, VoiceCraft \cite{peng2024voicecraft}), while non-autoregressive flow-matching and diffusion-based methods (e.g.,  E2-TTS \cite{eskimez2024e2}, Voicebox\cite{le2024voicebox}, F5-TTS \cite{chen2024f5}, P-flow \cite{kim2023p} and the Naturalspeech series \cite{shen2023naturalspeech,ju2024naturalspeech}) further improve quality and robustness. Hybrid frameworks such as CosyVoice combine token-level language modeling with flow-matching to enable speech infilling and strong zero-shot control~\cite{du2024cosyvoice}. 

% While these models deliver impressive naturalness and speaker similarity, they typically rely on conditional speaker modeling (embeddings or acoustic prompts) and large-scale training, which increases complexity, compute, and data requirements that hinders reproducibility and accessibility.
Although these models achieve impressive naturalness and speaker similarity, they often depend on conditional speaker modeling (via embeddings or acoustic prompts) and large-scale training. This reliance increases model complexity, computational cost, and data requirements, ultimately hindering reproducibility and accessibility.
In contrast, we explore a simple, non-parametric alternative: SelectTTS performs speaker modeling via SSL feature–based frame selection from a target speaker’s reference audio, avoiding conditional speaker parameters and achieving competitive or superior speaker similarity with substantially lower complexity.

\subsection{UNIT SELECTION AND CONCATENATIVE TTS}
%Concatenative and unit-based selection based TTS was very popular in earlier development of TTS, due to their simplicity and robustness\cite{Taylor_2009}. These systems revolved around building mechanisms to stitch pre-recorded speech segments together to generate the desired speech content \cite{beutnagel1999t}. While this class of modeling could generate audio with high intelligibility and reproduce the original timbre of the voice actor \cite{hunt1996unit,black1998festival}, it eventually got replaced by predictive and generative models.
%Consequently, with the success of deep learning methods, these earlier methods seem to be abandoned. In this work, we borrow key principles applied in unit-selection based TTS and extend its capabilities with modern modeling paradigms.

Concatenative and unit-selection based TTS was once the dominant paradigm in speech synthesis, valued for its simplicity, robustness, and ability to preserve the natural timbre of recorded speakers \cite{Taylor_2009}. These systems operate by stitching together prerecorded speech segments to form the desired utterance \cite{beutnagel1999t,hunt1996unit,black1998festival}. Although intelligible and natural-sounding, they were eventually replaced by predictive and generative neural models. We believe that the principle of directly reusing recorded speech fragments remains compelling for speaker similarity and efficiency. In this work, we revisit unit selection in the modern context: by leveraging self-supervised learning (SSL) features, we extend the classic idea of frame and unit selection to enable a low-complexity, frame-based TTS framework. This allows us to combine the strengths of unit-selection synthesis in preserving timbre with the superior modeling capabilities of neural models for semantic prediction and vocoding.
\vspace{-3mm}
\subsection{CONTINUOUS VS. DISCRETE SSL FEATURES FOR SPEECH SYNTHESIS}
Self-supervised learning (SSL) models such as WavLM \cite{chen2022wavlm} and HuBERT \cite{Hsu2021HuBERTSS} leverage large amounts of unlabeled data to learn rich speech representations. These features have been widely adopted in speech synthesis frameworks to define new training strategies and intermediate representations \cite{wang2023use}. Both continuous SSL embeddings and their discretized counterparts have found applications in TTS \cite{wang2025speech}. For instance, WavThruVec \cite{siuzdak2022wavthruvec} and HierSpeech++ \cite{lee2025hierspeech++} utilize wav2vec 2.0 continuous features \cite{baevski2020wav2vec} as intermediate representations between text and waveform, effectively enabling multi-stage modeling.  Continuous SSL features are high-dimensional and capture detailed acoustic and prosodic information, while discrete speech units provide efficient, linguistically grounded representations suitable for semantic-level tasks \cite{chang2024exploring,lee2024discrete}. In this work, we exploit both spaces: discrete SSL units for text-to-semantic prediction and frame selection, and continuous SSL features for vocoding. Our proposed subsequence matching and inverse k-means algorithms bridge these two domains, allowing SelectTTS to benefit from the strengths of each domain.

\subsection{SUMMARY OF RESEARCH GAP}
We highlight three key limitations in current multi-speaker TTS research:
\begin{itemize}
\item \textbf{High computational demands and limited reproducibility}: State-of-the-art systems rely on large model capacities and massive training datasets, making them costly to reproduce and restricting broader research participation. % this is PURELY about research 
\item \textbf{Dependence on speaker conditioning for generalization}: Most approaches jointly learn speech semantics and speaker conditioning, placing a heavy burden on the model to generalize to unseen speakers. This often reduces robustness in zero-shot scenarios.
% \item \textbf{Inefficiency for practical deployment}: Even when reproducible, current systems remain computationally heavy at training and inference, making them difficult to deploy in real-time or resource-constrained settings such as edge devices and low-resource languages. % this is about even if models are reproducible, they’re too heavy for real-world or resource-constrained deployment.

\item \textbf{Trade-off between performance, efficiency, and model complexity}: Current multi-speaker TTS systems face a significant trade-off, where achieving strong performance requires disproportionately high computational resources, limiting their practicality and accessibility.

\end{itemize}
To address these gaps, we propose a frame selection–based multi-speaker TTS framework for unseen speakers that balances high speaker similarity with low model complexity and practical reproducibility.

\section{THE PROPOSED METHOD}
The proposed SelectTTS framework consists of two training stages with one offline intermediate stage. In the first stage, we train a text-to-semantic-unit prediction model that generates discrete unit sequences from input text. Based on these predicted units, an offline frame selection step retrieves the corresponding frames from the target speaker’s reference speech, thereby recovering continuous SSL features from their discrete representations (Fig.~\ref{fig:frame_selection}). In the second stage, a vocoder is trained to synthesize the waveform from the selected frame-level SSL features (Fig.~\ref{fig:vocoding}). The following subsections describe each stage of the framework in detail.

 \subsection{SEMANTIC UNIT TOKENIZERS}
\label{sec:tokenize}
Frame selection in SelectTTS is performed in a discrete semantic-unit space, which provides a compact and linguistically meaningful representation of speech. These tokenizers serve as the bridge between text and speech, ensuring that semantic content is represented in a way that enables accurate matching to target-speaker frames. To this end, we define two tokenizers: one that discretizes continuous SSL features from speech, and another that predicts discrete units directly from text.  
\begin{figure}[htbp]
    \centering
    % First minipage (left side)
    \begin{minipage}[b]{0.24\textwidth}
        \centering
        \scalebox{0.92}{
        \includegraphics[width=0.8\textwidth]{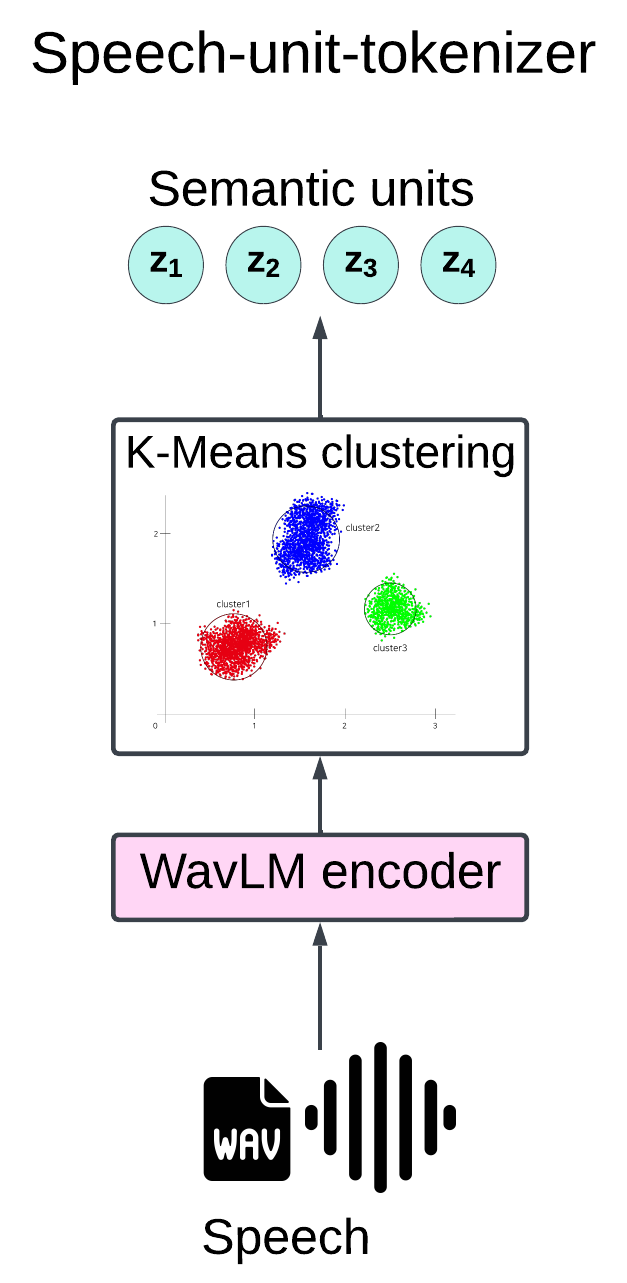}} % Replace with your figure file
        \caption{\textit{SpeechUnitTokenizer}: Discretization of SSL feature to speech semantic units}
        \label{fig:speechtokenizer}
    \end{minipage}
    % \hfill % this will add horizontal space between the minipages
    % Second minipage (right side)
    \begin{minipage}[b]{0.24\textwidth}
        \centering
        \scalebox{1.1}{
        \includegraphics[width=0.8\textwidth]{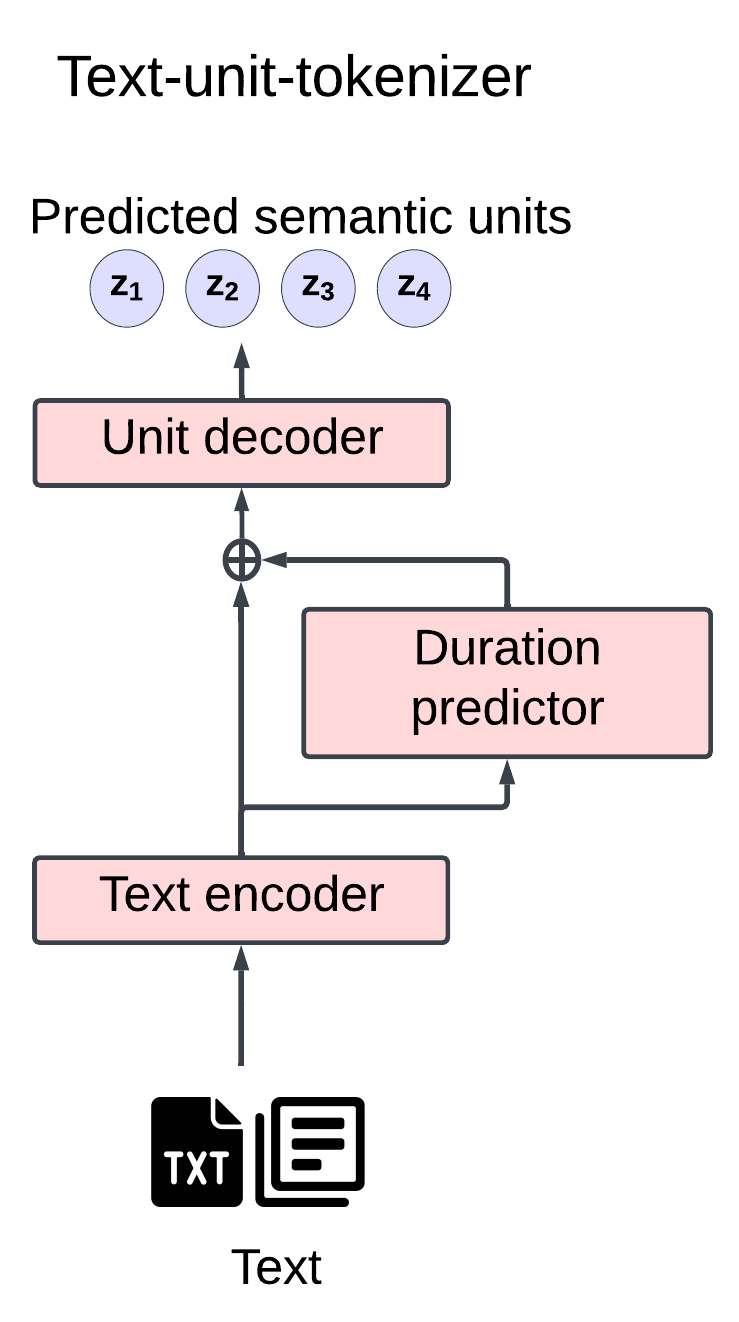} }% Replace with your figure file
        \caption{\textit{TextUnitTokenizer}: Prediction of speech semantic units from text}
        \label{fig:texttokenizer}
    \end{minipage}
    \vspace{-5mm}
\end{figure}

% We use the following tokenizers to bridge the modality gap between text and speech.
\vspace{-4mm}
\subsubsection{SPEECH-UNIT-TOKENIZER}
\label{speech_tokenize}
The speech-unit tokenizer converts speech into frame-level sequences of semantic units (Fig.~\ref{fig:speechtokenizer}). Continuous SSL features are first extracted at the frame level from a pre-trained SSL model. These features are then quantized using $k$-means clustering, yielding a discrete sequence of semantic units. We denote this transformation as $z = \textit{SpeechUnitTokenizer}(Z)$, where $Z$ is the continuous SSL feature sequence and $z$ is the corresponding discrete unit sequence.  

\vspace{-4mm}
\subsubsection{TEXT-UNIT-TOKENIZER}
\label{sec:texttokenize}
The text-unit tokenizer maps input text into frame-level sequences of discrete semantic units (Fig.~\ref{fig:texttokenizer}). We adopt a non-autoregressive FastSpeech2-based architecture \cite{ren2021fastspeech}, consisting of a text encoder, a duration prediction module, and a unit decoder. The model is trained on parallel text–speech data, where ground-truth speech units are obtained using the \textit{SpeechUnitTokenizer}. Phoneme durations are extracted using an external alignment tool \cite{mcauliffe2017montreal} and used as supervision for the duration predictor.  

During inference, the text-unit tokenizer takes phoneme sequences as input, predicts frame-level discrete semantic units, and upsamples them to match frame-level granularity based on the predicted durations. Predicting discrete units rather than continuous features simplifies the modeling task and reduces overall complexity. Furthermore, operating in the discrete space allows for more effective frame selection, enabling contiguous sequences of frames to be retrieved together and capturing inter-frame dependencies that would be difficult to preserve in the continuous space.

\vspace{-2mm}
\subsection{FRAME SELECTION PIPELINE}
\begin{figure*}[!ht] % The [t] option aligns the figure at the top of the page
    \centering
    \includegraphics[width=0.9\linewidth]{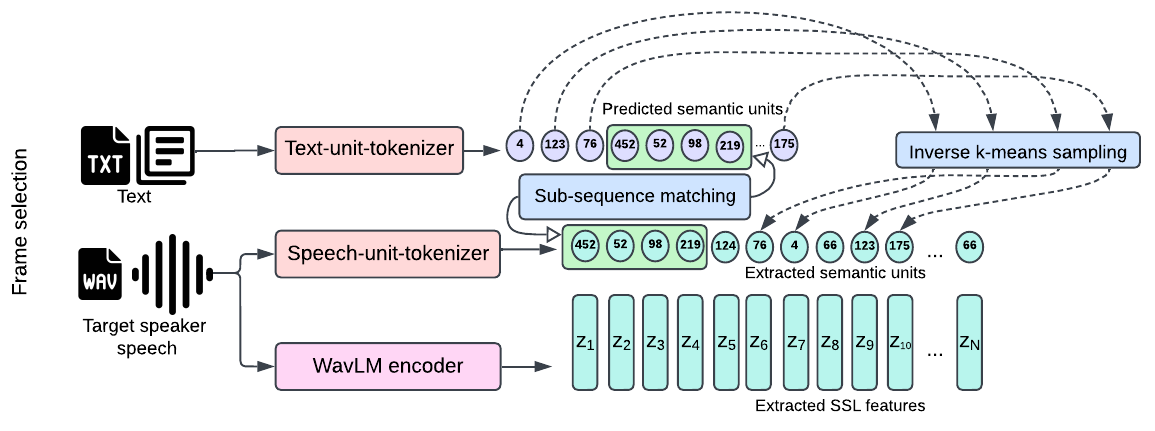}

  \caption{Proposed SelectTTS framework with the frame-selection method. In the frame selection, frames $z_1$,$z_2$,$z_3$,$z_4$ are chosen through subsequence matching and frames $z_7$, $z_9$,$z_6$ and $z_{10}$ are chosen via inverse k-means sampling. Here, red modules are trained online, pink modules are pre-trained and blue modules are offline non-parametric algorithms.}
  \label{fig:frame_selection}
\end{figure*}
\label{frame_selection}

\begin{figure*}[!ht] % The [t] option aligns the figure at the top of the page
    \centering
    \scalebox{0.8}{
    \includegraphics[width=0.8\linewidth]{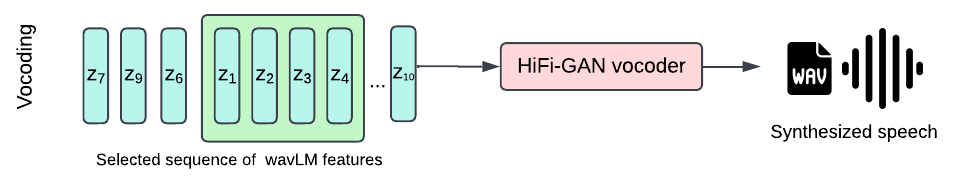}}

  \caption{Vocoding: The SSL feature sequence  is transformed to synthesized speech using the HiFi-GAN vocoder}
  \label{fig:vocoding}
\end{figure*}
We design a two-stage frame selection pipeline to map predicted semantic units to continuous SSL features of the target speaker. The pipeline first applies \textit{subsequence matching} to retrieve exact speech segments, and then falls back to \textit{inverse $k$-means sampling} for unmatched units. This combination balances segment-level prosody preservation with robust coverage.

\subsubsection{SUBSEQUENCE MATCHING FOR SEGMENT RETRIEVAL}
\label{Subsequence match}
%We formulate the sub-sequence matching problem as follows: 
%Let  $\hat{z}$ = \textit{TextUnitTokenizer(text)} represent a sequence of predicted semantic units obtained from the \textit{TextUnitTokenizer}. Similarly, let $z_{\text{ref}}$ = \textit{SpeechUnitTokenizer(speech)} denote the semantic unit sequence derived from the reference speech of the target speaker using the speech-unit tokenizer. 
%In our sub-sequence matching algorithm, we find sub-sequences of the predicted sequence $\hat{z}$ that is present in the reference unit sequence $z_{\text{ref}}$. Once we find these sub-sequences, we replace the sub-sequence-matched discrete unit sequence with the corresponding continuous SSL features from the reference speech as in Fig. \ref{fig:frame_selection}. We start from the longest sub-sequence length and iteratively replace the discrete units till there are no more matches. In the implementation, we begin by searching for matches with a maximum length of 10 to maintain computational efficiency, with a minimum length of 2. The intuition of sub-sequence matching is that by choosing chunks of speech segments at a time, we hope to get the most accurate match in the form of real speech segments instead of relying on only frame-level selection, to reduce artifacts and improve segment-level prosody. 

Let $\hat{z} = \textit{TextUnitTokenizer}(\text{text})$ denote the sequence of predicted semantic units from text, and $z_{\text{ref}} = \textit{SpeechUnitTokenizer}(\text{speech})$ the semantic unit sequence obtained from the target speaker’s reference speech.  The goal is to replace segments of $\hat{z}$ with their corresponding SSL features from $z_{\text{ref}}$ whenever possible.

Our algorithm proceeds as follows: starting from the longest possible subsequences (up to length 10, with a minimum length of 2), we iteratively search for subsequences of $\hat{z}$ contained in $z_{\text{ref}}$. When a match is found, the matched units are substituted with their continuous SSL features from the reference speech (see Fig.~\ref{fig:frame_selection}).  
By replacing contiguous segments rather than individual frames, this step preserves natural speech prosody and reduces artifacts compared to frame-level selection alone.

\vspace{-5mm}
\subsubsection{INVERSE $K$-MEANS SAMPLING FOR FALLBACK FRAME SELECTION}
%For frames without direct sub-sequence matches, we apply a strategy called inverse $k$-means sampling. In this approach, we identify frames from the reference speech that belong to the same discrete unit cluster as the predicted semantic unit. This effectively performs the reverse of the $k$-means clustering process by retrieving continuous SSL features from the discretized semantic space. If the predicted unit cluster is not represented in the reference speech - typically due to the limited reference duration - we instead select frames from the nearest nonempty cluster. Once the appropriate cluster is determined, the corresponding continuous SSL feature is obtained either by randomly sampling a frame from the target cluster or by averaging all SSL features associated with that cluster in the reference speech.

For units in $\hat{z}$ without direct subsequence matches, we turn to inverse $k$-means sampling. Each predicted unit is first assigned to its discrete cluster, as defined during the $SpeechUnitTokenizer$’s $k$-means training. We then retrieve continuous SSL features from the reference speech corresponding to that cluster.  If a cluster is absent in the reference (due to limited recording duration), the nearest non-empty cluster is used instead. Within the chosen cluster, the SSL feature can be obtained either by random frame sampling or by averaging all available features. This strategy effectively reverses the discretization process, ensuring coverage for all predicted units while maintaining consistency with the target speaker’s acoustic space.

\subsection{VOCODER}
The vocoder converts continuous WavLM features into audio waveforms, as illustrated in Fig.~\ref{fig:vocoding}. It is trained on paired ground-truth frame-level features and speech waveforms, independently of the frame selection process. However, this setup introduces a mismatch: during training the vocoder receives ground-truth features, while at inference it receives features generated by the frame selection algorithm. 

To overcome this mismatch, we incorporate frame selection during vocoder training. Specifically, given an input utterance, we treat every other utterance from the same speaker as potential reference speech. We extract ground-truth WavLM units from the input utterance using the \textit{SpeechUnitTokenizer}, and then perform frame selection as described in Section~\ref{frame_selection}. The resulting frame-level SSL features are used as input to the vocoder, with the target being the original ground-truth audio.  In preliminary experiments, we find that fine-tuning the vocoder on subsequence-matched features substantially reduces synthesis artifacts and improves audio quality.

\section{EXPERIMENTAL SETUP}
\subsection{SELECTTTS IMPLEMENTATION}
A key design choice in our framework is the selection of the SSL feature layer. We first experimented with layers 22 and 23 of WavLM-Large \cite{chen2022wavlm}, which are known to capture rich linguistic content and perform well on phoneme recognition tasks. However, we observed that features from these higher layers degraded speaker identity and prosody reproduction, both of which are crucial for multi-speaker TTS \cite{lin2023utility}. 
% Based on this finding, and consistent with prior work such as kNN-VC, we adopt features from layer 6 of WavLM-Large, which better preserve speaker-specific information, linguistic content information still being the most dominant thus suitable for semantic construction of synthesized speech.
Based on this finding, and in line with prior work such as kNN-VC, we adopt features from layer 6 of WavLM-Large. These features better preserve speaker-specific information while maintaining linguistic content as the dominant element, making them well-suited for the semantic construction of synthesized speech.
We next describe the three core modules of our implementation: the \textit{Speech-unit-tokenizer}, the \textit{Text-unit-tokenizer}, and the \textit{Vocoder}.
\setlength{\arrayrulewidth}{0.01mm}
\begin{table*}[ht]
\centering
\caption{Objective evaluation results (LibriTTS-R test-clean)
}
%\scalebox{0.86}
{\begin{tabular}{c@{}cccc}
\toprule

Method & WER(\%)$\downarrow$  &SECS $\uparrow$  & UTMOS $\uparrow$ \\
\noalign{\global\arrayrulewidth=0.05mm}\hline
\textit{Ground Truth} & \textit{3.55} & \textit{68.12} & \textit{4.22 } \\
\noalign{\global\arrayrulewidth=0.05mm}\hline
VALL-E      & 4.72 & 58.24  & 4.08  \\
XTTS-v2      &   4.23 & 60.26  & 4.16    \\

YourTTS      & 9.66& 50.83  & 3.61 \\

\noalign{\global\arrayrulewidth=0.05mm}\hline
SelectTTS (only inverse k-means (rand))      &   7.19& 62.84 & 3.46  \\
SelectTTS (only inverse k-means (avg))      &  6.49& 64.74   & 3.89  \\
SelectTTS (inverse k-means (rand) + subsequence-matching)      &  7.31& 61.57   & 3.99 \\
SelectTTS (inverse k-means (avg) + subsequence-matching)      & 6.67& 61.59  & 4.13  \\

\end{tabular}}

\label{table:CER}

\end{table*}

\begin{table*}[ht]
\centering
\caption{Objective evaluation results (VCTK)
}
% \scalebox{0.86}
{\begin{tabular}{c@{}cccc}
\toprule

Method & WER(\%)$\downarrow$  &SECS $\uparrow$  & UTMOS $\uparrow$ \\
\noalign{\global\arrayrulewidth=0.05mm}\hline
\textit{Ground Truth} & \textit{4.39} & \textit{69.77} & \textit{4.04} \\
\noalign{\global\arrayrulewidth=0.05mm}\hline
VALL-E  & 8.53   & 46.70 & 4.07 \\
XTTS-v2     &  4.89  & 54.12 & 4.00  \\

YourTTS  &  9.90  & 36.83 & 3.53 \\

\noalign{\global\arrayrulewidth=0.05mm}\hline
SelectTTS (only inverse k-means (avg)) & 7.18  & 60.61 & 3.69   \\
SelectTTS (inverse k-means (avg) + subsequence-matching)  & 7.01  & 54.13& 3.92  \\

\end{tabular}}

\label{table:VCTK}

\end{table*}
%This proved to be empirically strong, as our TTS models achieve low word error rates and high speaker similarity scores. 

\vspace{-6mm}
\subsubsection{SPEECH-UNIT-TOKENIZER}
The WavLM-Large encoder produces continuous feature vectors at 20ms intervals for 16kHz audio. We discretize these features using $k$-means clustering, as described in Section~\ref{speech_tokenize}. For our experiments, we use 2000 cluster centers. In preliminary trials with 100 and 500 clusters, we found that finer-grained quantization yielded improved intelligibility in the synthesized speech likely due to being less affected by the prediction errors, motivating our choice of 2000 clusters.

\vspace{-6mm}
\subsubsection{TEXT-UNIT-TOKENIZER}
We adapt the FastSpeech2 architecture \cite{ren2021fastspeech} to predict semantic units from text, replacing the standard Mel spectrogram prediction with semantic unit prediction. Ground-truth units are obtained by applying the \textit{SpeechUnitTokenizer} to the training audio. The network is optimized using the Adam optimizer with a learning rate of $5 \times 10^{-4}$ and a batch size corresponding to 10,000 phonemes. Training is performed on an NVIDIA RTX 3090 GPU, and the network typically converges within 20 minutes (around 10,000 steps). Our implementation builds upon the publicly available SpeechLM framework \cite{zhang2024speechlm}.

\vspace{-6mm}
\subsubsection{VOCODER} 
We use the HiFi-GAN V1 architecture \cite{kong2020hifi}, initialized from the pre-trained vocoder released with kNN-VC\footnote{\url{https://github.com/bshall/knn-vc}}. We fine-tune two vocoder variants using frames selected by our proposed algorithms: one trained solely with inverse $k$-means sampling, and another trained with a combination of subsequence matching and inverse $k$-means sampling.

% By doing this, we compensate for the shift in distribution caused by the frame selection algorithm\rsm{repetitive}.
%Our demo page features speech samples from both before and after the fine-tuning process.

\subsection{BASELINES}
We compare SelectTTS against three state-of-the-art multi-speaker TTS frameworks. \\
\textbf{YourTTS \cite{casanova2022yourtts}}:  
YourTTS is a zero-shot multi-speaker TTS framework built on top of VITS. It relies on speaker embeddings to capture speaker information for both seen and unseen speakers. For our experiments, we use the official implementation\footnote{\url{https://github.com/Edresson/YourTTS}} and the pre-trained model provided by COQUI\footnote{\url{https://github.com/coqui-ai/TTS}}. As a well-established multi-speaker model, YourTTS serves as a strong baseline conditioned on pre-trained speaker embeddings. \\
\textbf{XTTS-v2 \cite{casanova2024xtts}}:  
XTTS-v2 is a zero-shot TTS method built on top of the Tortoise model. It utilizes a perceiver-based conditioning encoder, decoder conditioning with pre-trained speaker embeddings, and a speaker consistency loss to better reproduce speaker characteristics. We use the official release from COQUI\footnote{\url{https://huggingface.co/coqui/XTTS-v2}} with default inference parameters. \\
\textbf{VALL-E \cite{chen2025neural}}:  
VALL-E is a language model–based TTS framework trained on over 45K hours of speech data. It first learns a neural codec language model and then performs TTS as a conditional language modeling task. Since there is no official release of VALL-E, we use Amphion's open-source implementation\footnote{\url{https://github.com/open-mmlab/Amphion/tree/main/egs/tts/VALLE_V2}}.

%which uses SpeechTokenizer \cite{zhang2024speechtokenizer} as the neural codec

% Specifically, we use the official releases of YourTTS\cite{casanova2022yourtts} and XTTS-v2\footnote{\url{https://huggingface.co/coqui/XTTS-v2}} \cite{casanova2024xtts}, both with default parameters. For VALL-E, we adopt the implementation provided in the Amphion toolkit\footnote{\url{https://github.com/open-mmlab/Amphion/tree/main/egs/tts/VALLE_V2}} \cite{chen2025neural}.

%\textbf{itemize them, talk more about them. also mention why these are the ideal baselines for our paper.}
% Since there is no official release of VALL-E \cite{chen2025neural}, we use Amphion's implementation\footnote{https://github.com/open-mmlab/Amphion/tree/main/egs/tts/VALLE\_V2}.
%which uses SpeechTokenizer \cite{zhang2024speechtokenizer} as the neural codec.

\subsection{TRAINING AND TEST DATA}
We train both the text-to-semantic-unit tokenizer and the HiFi-GAN vocoder on the LibriSpeech train-clean-100 dataset \cite{Panayotov2015LibrispeechAA}, which contains 100 hours of 16kHz read speech from 251 speakers. The LibriSpeech dev-clean subset is used for validation. 

For evaluation, we use two unseen test sets. From LibriTTS-R test-clean \cite{koizumi23_interspeech}, we select speakers not present in training and with at least 20 utterances (3–30 words each), yielding 31 speakers and 620 utterances in total. From VCTK \cite{vctk}, we sample 60 unseen speakers, each producing 10 randomly selected sentences, resulting in 600 utterances. To ensure comparability, we provide 5 minutes of reference audio for all variants of SelectTTS, as well as for YourTTS and XTTS-v2. Due to computational constraints, VALL-E is evaluated with 10–15 seconds of reference speech.

% \begin{table*}[th]
% \centering

% \caption{Subjective evaluations (95\% confidence interval)}
% %\scalebox{0.95}
% {\begin{tabular}{c c c}
% \toprule
% Method &
%  MOS$\uparrow$& SMOS$\uparrow$  \\
%                   \hline
                
% Ground Truth   & 4.20 $\pm$ 0.09 & -     \\
% \hline
% VALL-E    & 4.11 $\pm$ 0.08 & 3.62 $\pm$ 0.09\\
% XTTS-v2    & 4.10 $\pm$ 0.08 & 3.64 $\pm$  0.10\\
% YourTTS     & 3.15 $\pm$ 0.10 & 3.22 $\pm$ 0.09 \\
% \hline
% SelectTTS (inverse k-means (avg)) & 3.89 $\pm$ 0.08 & 3.71 $\pm$ 0.09   \\ 
% SelectTTS (inverse k-means (avg) + subsequence-matching) & 3.97 $\pm$ 0.08 & 3.72 $\pm$ 0.09 \\  \hline
% \end{tabular}}
% \label{table:sub}

% \end{table*}

% \begin{figure*}[!ht] % The [t] option aligns the figure at the top of the page
%     % \centering
%     \includegraphics[width=\linewidth]{figures/mos_smos_horizontal_legend_fixed5.pdf}

%   \caption{Subjective evaluations (95\% confidence interval) \rsm{Can you make two figures for MOS and SMOS separately. Methods would be side-by-side bars with different colors. It would be easier to compare.}}
%   \label{fig:MOS}
% \end{figure*}
\section{EXPERIMENTS AND RESULTS}
We evaluate four variants of the proposed SelectTTS framework.  
\begin{itemize}
    \item \textbf{only inverse k-means (rand):} inverse $k$-means sampling, where a frame is randomly selected from the predicted cluster.  
    \item \textbf{only inverse k-means (avg):} inverse $k$-means sampling, where the average of all frames in the predicted cluster is used for decoding.  
    \item \textbf{inverse k-means (rand) + subsequence-matching:} \\ extension of the first variant with subsequence matching.  
    \item \textbf{inverse k-means (avg) + subsequence-matching:} \\ extension of the second variant with subsequence matching.  
\end{itemize}

We note that the first two variants use the vocoder trained without subsequence-matching, whereas the last two variants use the vocoder trained with subsequence matching. This is done to match the training and inference for consistent vocoder behavior. Table~\ref{table:CER} reports objective evaluation results on the LibriTTS-R test set, and Table~\ref{table:VCTK} reports results on the VCTK test set.

%We evaluate four variants of the proposed SelectTTS framework. The first variant, inv-kmeans (rand), employs inverse $k$-means sampling by randomly selecting a frame from the predicted cluster. The second variant, inv-kmeans (avg), also uses inverse $k$-means sampling but decodes using the average of all frames within the predicted cluster. The third and fourth variants build on the first two by incorporating sub-sequence matching.

%Table~\ref{table:CER} presents objective evaluation results on the LibriTTS-R test set and table \ref{table:VCTK} presents the objective evaluation results on VCTK test set.

\subsection{OBJECTIVE EVALUATION OF INTELLIGIBILITY AND NATURALNESS}

\textbf{Intelligibility:} We measure intelligibility using the word error rate (WER) obtained from the Wav2Vec 2.0 Large automatic speech recognition (ASR) model\footnote{\url{https://huggingface.co/facebook/wav2vec2-large-960h-lv60-self}}. Results are consistent across both the LibriTTS-R and VCTK test sets: all variants of SelectTTS outperform YourTTS, and their performance approaches that of substantially larger SOTA models such as XTTS-v2 and VALL-E. While slightly higher error rates are expected given the significantly smaller training data used for SelectTTS, the achieved intelligibility is within an acceptable range for practical use. We notice that using the average of all frames in inverse k-means sampling improves intelligibility, probably due to the smoothing effect of averaging. \\ 
\textbf{Naturalness:} We assess naturalness using UTMOS~\cite{saeki2022utmos}, a MOS prediction system trained on human-annotated ratings. Most SelectTTS variants achieve higher UTMOS scores than YourTTS and perform comparably to SOTA models such as XTTS-v2 and VALL-E, with the exception of the variant using only inverse $k$-means (rand). Subsequence- matching yields notable gains in naturalness by selecting contiguous segments rather than isolated frames, enabling the reproduction of naturally occurring speech segments from the reference audio.  

These findings suggest:
\begin{itemize}
    \item Unit selection with SSL features can yield natural-sounding speech comparable to much larger SOTA models.  
    \item Inverse $k$-means sampling alone provides reasonable naturalness, validating its effectiveness as a lightweight selection criterion.
    \item Subsequence matching consistently improves overall quality, highlighting its importance for enhancing the naturalness of SelectTTS.
\end{itemize}

\begin{table*}[!h]
\centering
\caption{Analysis of model complexity}
{\begin{tabular}{ccccc}
\toprule

Method & \#Parameters & Training data (hours) & RTF $\downarrow$ \\
                  \hline
VALL-E      &  594M& 45k & 1.325    \\      
XTTS-v2     &  466M&27k &  0.186  \\

YourTTS      &  86M& 474&   0.120   \\
\hline
SelectTTS (only inverse k-means (avg))   &  57M& 100& 0.290    \\
SelectTTS (inverse k-means (avg) + subsequence-matching)  &  57M& 100& 0.334    \\
\end{tabular}}

\label{tab:complexity}
% \label{table:objeval}

\end{table*}
\begin{table*}[!h]
\centering
\caption{Analysis of objective performance with varying reference speech lengths}
{\begin{tabular}{c|ccc|ccc}

%Method & Ref speech & CER(\%) $\downarrow$ & WER(\%)$\downarrow$      & EER(\%)$\uparrow$ &SECS $\uparrow$ \\
 & \multicolumn{3}{c|}{WER(\%)$\downarrow$} & \multicolumn{3}{c}{SECS $\uparrow$} \\
\hline
Method & 3min & 1min & 30s  & 3min & 1min & 30s  \\                  
                  \hline
                
SelectTTS (only inverse k-means (avg))    &  6.55& 8.86& 13.35 & 63.85 & 62.17 &58.62   \\
SelectTTS (inverse k-means (avg) + subsequence-matching)   &  6.99& 8.87& 11.73 & 60.96 & 59.96 & 57.34  \\
\end{tabular}}
\label{table:refdur}
\end{table*}
 \vspace{-3mm}
\subsection{OBJECTIVE EVALUATION OF SPEAKER SIMILARITY}
We evaluate speaker similarity across models using 5 minutes of reference audio for all systems to ensure fairness. Speaker similarity is quantified using the speaker embedding cosine similarity (SECS) \cite{gibiansky2017deep}, computed between the ground-truth audio and the synthesized utterances. We extract embeddings using the ECAPA-TDNN model~\cite{desplanques20_interspeech}.

Our results show that SelectTTS consistently and significantly outperforms all state-of-the-art baselines in terms of objective speaker similarity. We attribute this improvement to three main factors:
\begin{itemize}
    \item Baselines predict speaker traits, which introduces inevitable prediction errors.
    \item SelectTTS directly reuses frames from the reference audio, similar to unit-selection TTS, thereby preserving speaker characteristics.  
    \item The SSL features used to construct frame sequences retain rich speaker-specific information, further enhancing acoustic similarity to the target speaker.
\end{itemize}
Among SelectTTS variants, inverse $k$-means sampling with averaging achieves the highest speaker similarity. This outcome is expected, as it leverages all frames within a target unit cluster to reconstruct the speaker’s voice thus utilizing more speaker-specific information, while subsequence matching operates on sequences of individual frames. These findings highlight the potential of SelectTTS: its performance can be further enhanced by designing strategies that exploit additional information from the reference speech.
\vspace{-4mm}
\subsection{SUBJECTIVE EVALUATION OF NATURALNESS AND SPEAKER SIMILARITY}
We conduct listening experiments with 21 participants to assess the naturalness and speaker similarity of synthesized speech. For evaluation, we use 120 utterances from six unseen speakers (three male and three female) in the LibriTTS-R test set. 

\textbf{Naturalness:} In Mean Opinion Score (MOS) test \cite{casanova2022yourtts}, participants rate the naturalness of speech samples on a 5-point scale. Each participant evaluates 12 utterances per method. As reported in Figure 5, SelectTTS achieves an MOS score close to 4, comparable to strong baselines, with subsequence matching further improving perceived naturalness. These results confirm that our low-complexity framework can reach a naturalness level on par with much larger systems. 

\textbf{Speaker similarity:} We also conduct a Speaker Mean Opinion Score (SMOS) test \cite{casanova2022yourtts}, in which participants are presented with a ground-truth recording and a synthesized utterance, and asked to rate their similarity on a 5-point scale. SelectTTS obtains strong SMOS scores (as shown in Fig.~\ref{fig:smos}), indicating speaker resemblance on par with—or exceeding—that of larger baseline models. This demonstrates that SelectTTS effectively preserves speaker characteristics even in a zero-shot setting, despite being substantially simpler and lighter-weight than competing approaches.

Overall, subjective evaluations validate that SelectTTS achieves both naturalness and speaker similarity competitive with state-of-the-art models, while maintaining much lower complexity.
 \begin{figure}[h]
  \centering
  \begin{minipage}[t]{0.40\textwidth}
    \centering
    \includegraphics[width=\linewidth]{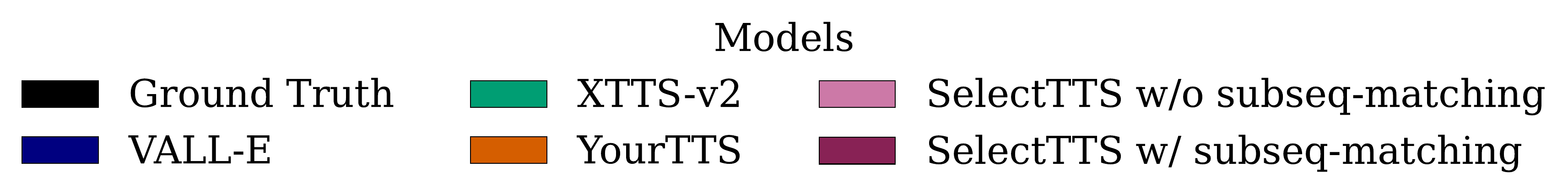}
    % \caption{Speaker Similarity MOS (SMOS).}
    \label{fig:smos}
  \end{minipage}
  % Left figure
  \begin{minipage}[t]{0.45\textwidth}
    \centering
    \includegraphics[width=\linewidth]{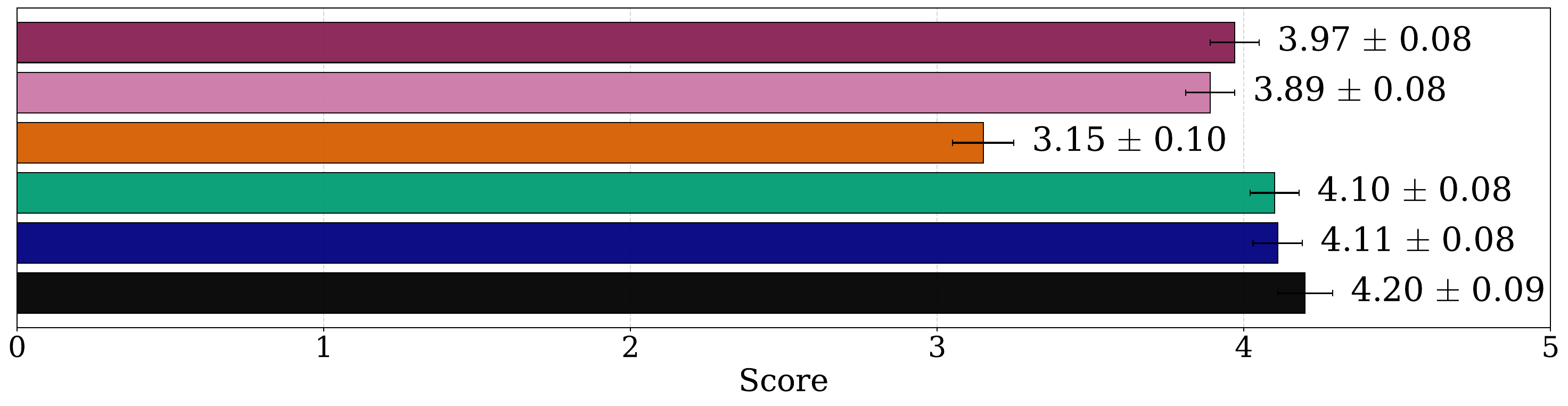}
    \vspace{-5mm}
    \caption{Mean Opinion Score (MOS)}
    \label{fig:mos}
  \end{minipage}\vfill
  %
  % Right figure
  \vspace{5mm}
  \begin{minipage}[t]{0.45\textwidth}
    \centering
    \includegraphics[width=\linewidth]{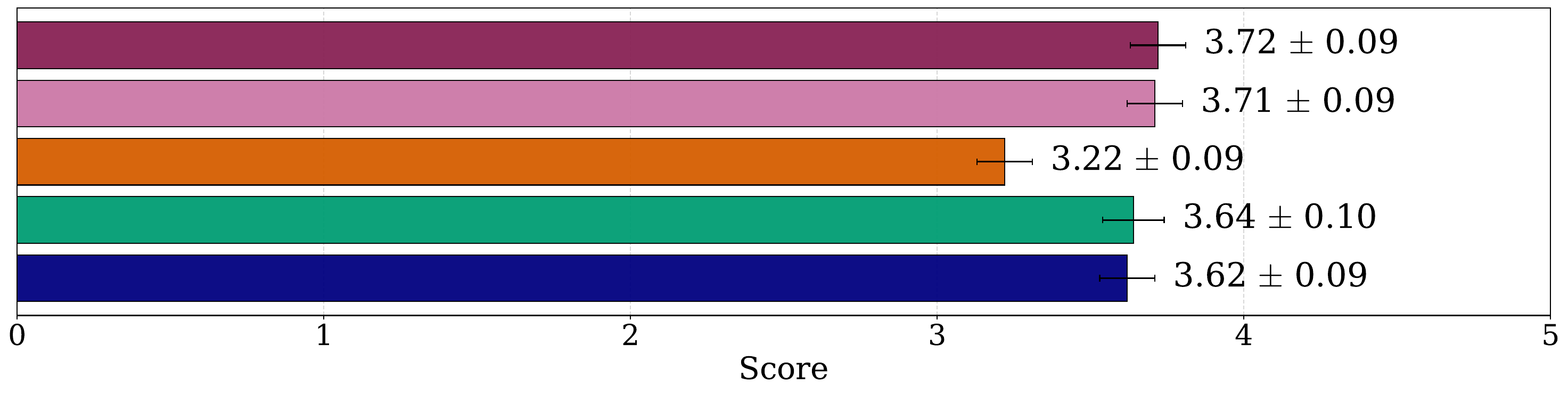}
    \vspace{-4mm}
    \caption{Speaker Similarity MOS (SMOS)}
    \label{fig:smos}
  \end{minipage}
% \vspace{-5mm}
\end{figure}
\vspace{-4mm}
\subsection{MODEL COMPLEXITY}
The previous experiments on intelligibility, naturalness, and speaker similarity, evaluated through both objective metrics and subjective listening tests, demonstrated that SelectTTS achieves performance comparable to strong state-of-the-art baselines. We now turn to model complexity, an equally important factor for the accessibility and practicality of multi-speaker TTS systems.

We compare models along three dimensions: number of parameters, scale of training data, and real-time factor (RTF). As summarized in Table~\ref{tab:complexity}, SelectTTS requires $8\times$ fewer parameters than XTTS-v2 and $10\times$ fewer than VALL-E, while using $270\times$ and $450\times$ less training data, respectively. In terms of RTF, SelectTTS substantially outperforms the autoregressive language-model-based TTS framework VALL-E, and is comparable to XTTS-v2 and YourTTS. We note that the offline frame selection approach does not utilize GPU. We anticipate that further optimizations of the frame selection algorithms could narrow the remaining RTF gap.

 \vspace{-2mm}
\subsection{PERFORMANCE ANALYSIS WITH CHANGE IN REFERENCE SPEECH DURATION}
The previous results established that SelectTTS achieves competitive performance and operates with substantially lower complexity than state-of-the-art baselines. We now analyze its robustness to changes in reference speech duration, since SelectTTS relies on selecting frames from the reference audio. We measure WER and SECS scores with three different amounts of reference audio: 3 minutes, 1 minute, and 30 seconds. Table~\ref{table:refdur} summarizes the results. Although performance gradually decreases with shorter references, SelectTTS remains competitive even with only 30 seconds of reference speech, demonstrating that the framework is effective even under limited reference conditions.

\vspace{-4mm}
\section{DISCUSSION}
\subsection{THE ROLE OF UNIT SELECTION}
Our experiments with different unit selection strategies (e.g., inverse $k$-means vs. subsequence matching) demonstrate that the proposed framework is highly customizable and open to further development. Simple cluster averaging during inverse $k$-means sampling yields substantial gains in intelligibility and speaker similarity compared to random sampling, while subsequence matching consistently enhances naturalness. Moreover, the clear separation between text-to-semantic prediction and speaker modeling allows these modules to be independently improved or replaced, providing flexibility for future research. We see many promising directions to further refine unit selection with more advanced strategies. Just as unit-selection–based TTS once played a central role in speech synthesis, our proposed modernized framework has the potential to revive this paradigm as a pathway toward efficient and lightweight zero-shot TTS.

\subsection{RELATION TO KNN-VC}
SelectTTS is inspired by the retrieval-based voice conversion method kNN-VC \cite{baas23_interspeech}, but it introduces a fundamentally different task and framework. While kNN-VC directly retrieves SSL features from input speech for conversion, SelectTTS predicts semantic representations from text and performs frame selection from reference speech to synthesize new utterances. This design bridges text and speaker-specific SSL features, enabling low-complexity TTS for unseen speakers. 

\subsection{FUTURE WORK: LOW-RESOURCE AND REAL-TIME APPLICATIONS OF SELECTTTS}
Beyond English benchmarks, SelectTTS has potential for deployment in low-resource languages and real-time scenarios. Its modular design allows the semantic prediction task—trainable on small speech-text datasets from a target low-resource language—to be decoupled from the acoustic modeling task. This modular separation is particularly advantageous for languages with limited corpora and may serve as a catalyst for breakthroughs in accessibility. Moreover, the low model complexity and competitive real-time factor suggest feasibility for on-device or streaming applications, without sacrificing synthesis quality. Future work will explore adaptation strategies for multilingual settings and further optimizations of frame selection to reduce latency.

% Beyond English benchmarks, SelectTTS demonstrates strong potential for deployment in low-resource and real-time settings. Its modular design allows the semantic prediction task—trainable on small speech-text datasets from a target low-resource language—to be decoupled from the acoustic modeling task. Prior work (kNN-VC) has shown that SSL representations trained on high-resource languages can effectively reconstruct speech in previously unseen languages. This modular separation is particularly advantageous for languages with limited corpora and may serve as a catalyst for breakthroughs in accessibility. In addition, SelectTTS delivers state-of-the-art zero-shot TTS performance with relatively low model complexity, positioning it as a promising candidate for on-device applications without sacrificing synthesis quality. Future directions include investigating adaptation strategies for multilingual settings and further optimizing frame selection to reduce inference latency.

%this is sth we mention in intro, related work. i think it would be ideal to state the link again under discussion and mention that this is our future work
\section{CONCLUSION}
%\subsection{Effect of number of discrete units}
In this paper, we propose SelectTTS, a multi-speaker TTS framework with lower model complexity that synthesizes high-quality speech closely resembling the target speaker's voice by directly utilizing frames from unseen speakers. We demonstrate that combining semantic unit-based frame selection with vocoding over SSL feature sequences provides a simple yet effective approach for modeling unseen speakers, achieving speaker similarity performance on-par with state-of-the-art TTS systems. SelectTTS significantly reduces model complexity and data requirements, opening possibilities for extending state-of-the-art multi-speaker TTS to low-resource settings and languages.
% \vspace{-2mm}
\bibliographystyle{IEEEtran}
\bibliography{mybib}

\begin{thebibliography}{10}
\providecommand{\url}[1]{#1}
\csname url@samestyle\endcsname
\providecommand{\newblock}{\relax}
\providecommand{\bibinfo}[2]{#2}
\providecommand{\BIBentrySTDinterwordspacing}{\spaceskip=0pt\relax}
\providecommand{\BIBentryALTinterwordstretchfactor}{4}
\providecommand{\BIBentryALTinterwordspacing}{\spaceskip=\fontdimen2\font plus
\BIBentryALTinterwordstretchfactor\fontdimen3\font minus \fontdimen4\font\relax}
\providecommand{\BIBforeignlanguage}[2]{{%
\expandafter\ifx\csname l@#1\endcsname\relax
\typeout{** WARNING: IEEEtran.bst: No hyphenation pattern has been}%
\typeout{** loaded for the language `#1'. Using the pattern for}%
\typeout{** the default language instead.}%
\else
\language=\csname l@#1\endcsname
\fi
#2}}
\providecommand{\BIBdecl}{\relax}
\BIBdecl

\bibitem{le2024voicebox}
M.~Le, A.~Vyas, B.~Shi, B.~Karrer, L.~Sari, R.~Moritz, M.~Williamson, V.~Manohar, Y.~Adi, J.~Mahadeokar \emph{et~al.}, ``Voicebox: Text-guided multilingual universal speech generation at scale,'' \emph{Advances in neural information processing systems}, vol.~36, 2024.

\bibitem{chen2025neural}
S.~Chen, C.~Wang, Y.~Wu, Z.~Zhang, L.~Zhou, S.~Liu, Z.~Chen, Y.~Liu, H.~Wang, J.~Li \emph{et~al.}, ``Neural codec language models are zero-shot text to speech synthesizers,'' \emph{IEEE Transactions on Audio, Speech and Language Processing}, 2025.

\bibitem{casanova2024xtts}
E.~Casanova, K.~Davis, E.~G{\"o}lge, G.~G{\"o}knar, I.~Gulea, L.~Hart, A.~Aljafari, J.~Meyer, R.~Morais, S.~Olayemi \emph{et~al.}, ``Xtts: a massively multilingual zero-shot text-to-speech model,'' in \emph{Proc. Interspeech 2024}, 2024, pp. 4978--4982.

\bibitem{lajszczak2024base}
M.~{\L}ajszczak, G.~C{\'a}mbara, Y.~Li, F.~Beyhan, A.~van Korlaar, F.~Yang, A.~Joly, {\'A}.~Mart{\'\i}n-Cortinas, A.~Abbas, A.~Michalski \emph{et~al.}, ``Base tts: Lessons from building a billion-parameter text-to-speech model on 100k hours of data,'' \emph{arXiv preprint arXiv:2402.08093}, 2024.

\bibitem{chen20r_interspeech}
M.~Chen, X.~Tan, Y.~Ren, J.~Xu, H.~Sun, S.~Zhao, and T.~Qin, ``{MultiSpeech: Multi-Speaker Text to Speech with Transformer},'' in \emph{Proc. Interspeech 2020}, 2020, pp. 4024--4028.

\bibitem{jia2018transfer}
Y.~Jia, Y.~Zhang, R.~Weiss, Q.~Wang, J.~Shen, F.~Ren, P.~Nguyen, R.~Pang, I.~Lopez~Moreno, Y.~Wu \emph{et~al.}, ``Transfer learning from speaker verification to multispeaker text-to-speech synthesis,'' \emph{Advances in neural information processing systems}, vol.~31, 2018.

\bibitem{casanova2022yourtts}
E.~Casanova, J.~Weber, C.~D. Shulby, A.~C. Junior, E.~G{\"o}lge, and M.~A. Ponti, ``Yourtts: Towards zero-shot multi-speaker tts and zero-shot voice conversion for everyone,'' in \emph{International Conference on Machine Learning}.\hskip 1em plus 0.5em minus 0.4em\relax PMLR, 2022, pp. 2709--2720.

\bibitem{melechovsky2024dart}
J.~Melechovsky, A.~Mehrish, B.~SISMAN, and D.~Herremans, ``Dart: Disentanglement of accent and speaker representation in multispeaker text-to-speech,'' in \emph{Audio Imagination: NeurIPS 2024 Workshop AI-Driven Speech, Music, and Sound Generation}.

\bibitem{gibiansky2017deep}
A.~Gibiansky, S.~Arik, G.~Diamos, J.~Miller, K.~Peng, W.~Ping, J.~Raiman, and Y.~Zhou, ``Deep voice 2: Multi-speaker neural text-to-speech,'' \emph{Advances in neural information processing systems}, vol.~30, 2017.

\bibitem{cooper2020zero}
E.~Cooper, C.-I. Lai, Y.~Yasuda, F.~Fang, X.~Wang, N.~Chen, and J.~Yamagishi, ``Zero-shot multi-speaker text-to-speech with state-of-the-art neural speaker embeddings,'' in \emph{ICASSP 2020-2020 IEEE International Conference on Acoustics, Speech and Signal Processing (ICASSP)}.\hskip 1em plus 0.5em minus 0.4em\relax IEEE, 2020, pp. 6184--6188.

\bibitem{jawaid2024style}
A.~Jawaid, S.~S. Chandra, J.~Lu, and B.~SISMAN, ``Style mixture of experts for expressive text-to-speech synthesis,'' in \emph{Audio Imagination: NeurIPS 2024 Workshop AI-Driven Speech, Music, and Sound Generation}.

\bibitem{xin2024rall}
D.~Xin, X.~Tan, K.~Shen, Z.~Ju, D.~Yang, Y.~Wang, S.~Takamichi, H.~Saruwatari, S.~Liu, J.~Li \emph{et~al.}, ``Rall-e: Robust codec language modeling with chain-of-thought prompting for text-to-speech synthesis,'' \emph{arXiv preprint arXiv:2404.03204}, 2024.

\bibitem{peng2024voicecraft}
\BIBentryALTinterwordspacing
P.~Peng, P.-Y. Huang, S.-W. Li, A.~Mohamed, and D.~Harwath, ``{V}oice{C}raft: Zero-shot speech editing and text-to-speech in the wild,'' in \emph{Proceedings of the 62nd Annual Meeting of the Association for Computational Linguistics (Volume 1: Long Papers)}, Aug. 2024, pp. 12\,442--12\,462. [Online]. Available: \url{https://aclanthology.org/2024.acl-long.673/}
\BIBentrySTDinterwordspacing

\bibitem{du2024cosyvoice}
Z.~Du, Q.~Chen, S.~Zhang, K.~Hu, H.~Lu, Y.~Yang, H.~Hu, S.~Zheng, Y.~Gu, Z.~Ma \emph{et~al.}, ``Cosyvoice: A scalable multilingual zero-shot text-to-speech synthesizer based on supervised semantic tokens,'' \emph{arXiv preprint arXiv:2407.05407}, 2024.

\bibitem{wang2024speechx}
X.~Wang, M.~Thakker, Z.~Chen, N.~Kanda, S.~E. Eskimez, S.~Chen, M.~Tang, S.~Liu, J.~Li, and T.~Yoshioka, ``Speechx: Neural codec language model as a versatile speech transformer,'' \emph{IEEE/ACM Transactions on Audio, Speech, and Language Processing}, 2024.

\bibitem{baevski2020wav2vec}
A.~Baevski, Y.~Zhou, A.~Mohamed, and M.~Auli, ``wav2vec 2.0: A framework for self-supervised learning of speech representations,'' \emph{Advances in neural information processing systems}, vol.~33, pp. 12\,449--12\,460, 2020.

\bibitem{chen2022wavlm}
S.~Chen, C.~Wang, Z.~Chen, Y.~Wu, S.~Liu, Z.~Chen, J.~Li, N.~Kanda, T.~Yoshioka, X.~Xiao \emph{et~al.}, ``Wavlm: Large-scale self-supervised pre-training for full stack speech processing,'' \emph{IEEE Journal of Selected Topics in Signal Processing}, vol.~16, no.~6, pp. 1505--1518, 2022.

\bibitem{pasad2021layer}
A.~Pasad, J.-C. Chou, and K.~Livescu, ``Layer-wise analysis of a self-supervised speech representation model,'' in \emph{2021 IEEE Automatic Speech Recognition and Understanding Workshop (ASRU)}.\hskip 1em plus 0.5em minus 0.4em\relax IEEE, 2021, pp. 914--921.

\bibitem{hsu2021hubert}
W.-N. Hsu, B.~Bolte, Y.-H.~H. Tsai, K.~Lakhotia, R.~Salakhutdinov, and A.~Mohamed, ``Hubert: Self-supervised speech representation learning by masked prediction of hidden units,'' \emph{IEEE/ACM transactions on audio, speech, and language processing}, vol.~29, pp. 3451--3460, 2021.

\bibitem{baas23_interspeech}
M.~Baas, B.~{van Niekerk}, and H.~Kamper, ``{Voice Conversion With Just Nearest Neighbors},'' in \emph{Proc. INTERSPEECH 2023}, 2023, pp. 2053--2057.

\bibitem{casanova2021sc}
E.~Casanova, C.~Shulby, E.~Gölge, N.~M. Müller, F.~S. de~Oliveira, A.~{Candido Jr.}, A.~da~Silva~Soares, S.~M. Aluisio, and M.~A. Ponti, ``Sc-glowtts: An efficient zero-shot multi-speaker text-to-speech model,'' in \emph{Interspeech 2021}, 2021, pp. 3645--3649.

\bibitem{huang2022meta}
S.-F. Huang, C.-J. Lin, D.-R. Liu, Y.-C. Chen, and H.-y. Lee, ``Meta-tts: Meta-learning for few-shot speaker adaptive text-to-speech,'' \emph{IEEE/ACM Transactions on Audio, Speech, and Language Processing}, vol.~30, pp. 1558--1571, 2022.

\bibitem{kim23k_interspeech}
H.~Kim, S.~Kim, J.~Yeom, and S.~Yoon, ``{UnitSpeech: Speaker-adaptive Speech Synthesis with Untranscribed Data},'' in \emph{Proc. INTERSPEECH 2023}, 2023, pp. 3038--3042.

\bibitem{lee2022hierspeech}
S.-H. Lee, S.-B. Kim, J.-H. Lee, E.~Song, M.-J. Hwang, and S.-W. Lee, ``Hierspeech: Bridging the gap between text and speech by hierarchical variational inference using self-supervised representations for speech synthesis,'' \emph{Advances in Neural Information Processing Systems}, vol.~35, pp. 16\,624--16\,636, 2022.

\bibitem{eskimez2024e2}
S.~E. Eskimez, X.~Wang, M.~Thakker, C.~Li, C.-H. Tsai, Z.~Xiao, H.~Yang, Z.~Zhu, M.~Tang, X.~Tan \emph{et~al.}, ``E2 tts: Embarrassingly easy fully non-autoregressive zero-shot tts,'' in \emph{2024 IEEE Spoken Language Technology Workshop (SLT)}.\hskip 1em plus 0.5em minus 0.4em\relax IEEE, 2024, pp. 682--689.

\bibitem{chen2024f5}
Y.~Chen, Z.~Niu, Z.~Ma, K.~Deng, C.~Wang, J.~Zhao, K.~Yu, and X.~Chen, ``F5-tts: A fairytaler that fakes fluent and faithful speech with flow matching,'' \emph{arXiv preprint arXiv:2410.06885}, 2024.

\bibitem{kim2023p}
S.~Kim, K.~Shih, J.~F. Santos, E.~Bakhturina, M.~Desta, R.~Valle, S.~Yoon, B.~Catanzaro \emph{et~al.}, ``P-flow: A fast and data-efficient zero-shot tts through speech prompting,'' \emph{Advances in Neural Information Processing Systems}, vol.~36, pp. 74\,213--74\,228, 2023.

\bibitem{shen2023naturalspeech}
\BIBentryALTinterwordspacing
K.~Shen, Z.~Ju, X.~Tan, E.~Liu, Y.~Leng, L.~He, T.~Qin, sheng zhao, and J.~Bian, ``Naturalspeech 2: Latent diffusion models are natural and zero-shot speech and singing synthesizers,'' in \emph{The Twelfth International Conference on Learning Representations}, 2024. [Online]. Available: \url{https://openreview.net/forum?id=Rc7dAwVL3v}
\BIBentrySTDinterwordspacing

\bibitem{ju2024naturalspeech}
Z.~Ju, Y.~Wang, K.~Shen, X.~Tan, D.~Xin, D.~Yang, E.~Liu, Y.~Leng, K.~Song, S.~Tang, Z.~Wu, T.~Qin, X.~Li, W.~Ye, S.~Zhang, J.~Bian, L.~He, J.~Li, and S.~Zhao, ``Naturalspeech 3: Zero-shot speech synthesis with factorized codec and diffusion models,'' in \emph{ICML}, ser. ICML'24, 2024.

\bibitem{Taylor_2009}
P.~Taylor, \emph{Unit-selection synthesis}.\hskip 1em plus 0.5em minus 0.4em\relax Cambridge University Press, 2009, p. 474–516.

\bibitem{beutnagel1999t}
M.~Beutnagel, A.~Conkie, J.~Schroeter, Y.~Stylianou, and A.~Syrdal, ``The at\&t next-gen tts system,'' in \emph{Joint meeting of ASA, EAA, and DAGA}, vol.~15.\hskip 1em plus 0.5em minus 0.4em\relax Berlin, Germany, 1999, pp. 18--24.

\bibitem{hunt1996unit}
A.~J. Hunt and A.~W. Black, ``Unit selection in a concatenative speech synthesis system using a large speech database,'' in \emph{1996 IEEE international conference on acoustics, speech, and signal processing conference proceedings}, vol.~1.\hskip 1em plus 0.5em minus 0.4em\relax IEEE, 1996, pp. 373--376.

\bibitem{black1998festival}
A.~Black, P.~Taylor, R.~Caley, and R.~Clark, ``The festival speech synthesis system,'' 1998.

\bibitem{Hsu2021HuBERTSS}
\BIBentryALTinterwordspacing
W.-N. Hsu, B.~Bolte, Y.-H.~H. Tsai, K.~Lakhotia, R.~Salakhutdinov, and A.~rahman Mohamed, ``Hubert: Self-supervised speech representation learning by masked prediction of hidden units,'' \emph{IEEE/ACM Transactions on Audio, Speech, and Language Processing}, vol.~29, pp. 3451--3460, 2021. [Online]. Available: \url{https://api.semanticscholar.org/CorpusID:235421619}
\BIBentrySTDinterwordspacing

\bibitem{wang2023use}
S.~Wang, G.~E. Henter, J.~Gustafson, and E.~Szekely, ``On the use of self-supervised speech representations in spontaneous speech synthesis,'' in \emph{Proc. SSW 2023}, 2023, pp. 163--169.

\bibitem{wang2025speech}
D.~Wang, J.~Li, M.~Cui, D.~Yang, X.~Chen, and H.~Meng, ``Speech discrete tokens or continuous features? a comparative analysis for spoken language understanding in speechllms,'' \emph{arXiv preprint arXiv:2508.17863}, 2025.

\bibitem{siuzdak2022wavthruvec}
H.~Siuzdak, P.~Dura, P.~van Rijn, and N.~Jacoby, ``Wavthruvec: Latent speech representation as intermediate features for neural speech synthesis,'' in \emph{Interspeech 2022}, 2022, pp. 833--837.

\bibitem{lee2025hierspeech++}
S.-H. Lee, H.-Y. Choi, S.-B. Kim, and S.-W. Lee, ``Hierspeech++: Bridging the gap between semantic and acoustic representation of speech by hierarchical variational inference for zero-shot speech synthesis,'' \emph{IEEE Transactions on Neural Networks and Learning Systems}, 2025.

\bibitem{chang2024exploring}
X.~Chang, B.~Yan, K.~Choi, J.-W. Jung, Y.~Lu, S.~Maiti, R.~Sharma, J.~Shi, J.~Tian, S.~Watanabe \emph{et~al.}, ``Exploring speech recognition, translation, and understanding with discrete speech units: A comparative study,'' in \emph{ICASSP 2024-2024 IEEE International Conference on Acoustics, Speech and Signal Processing (ICASSP)}.\hskip 1em plus 0.5em minus 0.4em\relax IEEE, 2024, pp. 11\,481--11\,485.

\bibitem{lee2024discrete}
P.~H. Lee, I.~R. Ulgen, and B.~Sisman, ``Discrete unit based masking for improving disentanglement in voice conversion,'' in \emph{2024 IEEE Spoken Language Technology Workshop (SLT)}.\hskip 1em plus 0.5em minus 0.4em\relax IEEE, 2024, pp. 742--749.

\bibitem{ren2021fastspeech}
\BIBentryALTinterwordspacing
Y.~Ren, C.~Hu, X.~Tan, T.~Qin, S.~Zhao, Z.~Zhao, and T.-Y. Liu, ``Fastspeech 2: Fast and high-quality end-to-end text to speech,'' in \emph{International Conference on Learning Representations}, 2021. [Online]. Available: \url{https://openreview.net/forum?id=piLPYqxtWuA}
\BIBentrySTDinterwordspacing

\bibitem{mcauliffe2017montreal}
M.~McAuliffe, M.~Socolof, S.~Mihuc, M.~Wagner, and M.~Sonderegger, ``Montreal forced aligner: Trainable text-speech alignment using kaldi.'' in \emph{Interspeech}, vol. 2017, 2017, pp. 498--502.

\bibitem{lin2023utility}
G.-T. Lin, C.-L. Feng, W.-P. Huang, Y.~Tseng, T.-H. Lin, C.-A. Li, H.-y. Lee, and N.~G. Ward, ``On the utility of self-supervised models for prosody-related tasks,'' in \emph{2022 IEEE Spoken Language Technology Workshop (SLT)}.\hskip 1em plus 0.5em minus 0.4em\relax IEEE, 2023, pp. 1104--1111.

\bibitem{zhang2024speechlm}
Z.~Zhang, S.~Chen, L.~Zhou, Y.~Wu, S.~Ren, S.~Liu, Z.~Yao, X.~Gong, L.~Dai, J.~Li \emph{et~al.}, ``Speechlm: Enhanced speech pre-training with unpaired textual data,'' \emph{IEEE/ACM Transactions on Audio, Speech, and Language Processing}, vol.~32, pp. 2177--2187, 2024.

\bibitem{kong2020hifi}
J.~Kong, J.~Kim, and J.~Bae, ``Hifi-gan: Generative adversarial networks for efficient and high fidelity speech synthesis,'' \emph{Advances in Neural Information Processing Systems}, vol.~33, pp. 17\,022--17\,033, 2020.

\bibitem{Panayotov2015LibrispeechAA}
\BIBentryALTinterwordspacing
V.~Panayotov, G.~Chen, D.~Povey, and S.~Khudanpur, ``Librispeech: An asr corpus based on public domain audio books,'' \emph{2015 IEEE International Conference on Acoustics, Speech and Signal Processing (ICASSP)}, pp. 5206--5210, 2015. [Online]. Available: \url{https://api.semanticscholar.org/CorpusID:2191379}
\BIBentrySTDinterwordspacing

\bibitem{koizumi23_interspeech}
Y.~Koizumi, H.~Zen, S.~Karita, Y.~Ding, K.~Yatabe, N.~Morioka, M.~Bacchiani, Y.~Zhang, W.~Han, and A.~Bapna, ``Libritts-r: A restored multi-speaker text-to-speech corpus,'' in \emph{Interspeech 2023}, 2023, pp. 5496--5500.

\bibitem{vctk}
J.~Yamagishi, C.~Veaux, and K.~MacDonald, ``{CSTR VCTK Corpus}: English multi-speaker corpus for {CSTR} voice cloning toolkit (version 0.92),'' 2019.

\bibitem{saeki2022utmos}
T.~Saeki, D.~Xin, W.~Nakata, T.~Koriyama, S.~Takamichi, and H.~Saruwatari, ``Utmos: Utokyo-sarulab system for voicemos challenge 2022,'' in \emph{Proc. Interspeech 2022}, 2022, pp. 4521--4525.

\bibitem{desplanques20_interspeech}
B.~Desplanques, J.~Thienpondt, and K.~Demuynck, ``{ECAPA-TDNN: Emphasized Channel Attention, Propagation and Aggregation in TDNN Based Speaker Verification},'' in \emph{Proc. Interspeech 2020}, 2020, pp. 3830--3834.

\end{thebibliography}

\end{document}